\documentclass[twocolumn,aps,showpacs,showkeys,floatfix]{revtex4}  

\usepackage{graphicx}
\usepackage{longtable}

\def\Var{\mathop{\mathrm{Var}}\nolimits}

\begin{document}

\author{Richard J. Mathar}
\email{mathar@strw.leidenuniv.nl}
\homepage{http://www.strw.leidenuniv.nl/~mathar}
\affiliation{
Leiden Observatory, Leiden University, P.O. Box 9513, 2300 RA Leiden, The Netherlands}

\thanks{
This work is supported by the NWO VICI grant
639.043.201
``Optical Interferometry: A new Method for Studies of Extrasolar Planets''
to A. Quirrenbach.
}

\date{\today}
\title{Computed Coupling Efficiencies of Kolmogorov Phase Screens into Single-Mode Optical Fibers}

\begin{abstract}
Coupling efficiencies of an electromagnetic field with a 
Kolmogorov phase statistics into a step-index fiber in its monomode
regime of wavelengths are computed
from the overlap integral  between the phase screens and the
far-field of the monomode at infrared wavelengths.

The phase screens are composed from Karhunen-Lo\`eve basis functions, 
optionally cutting off some of the eigenmodes of largest eigenvalue
as if Adaptive Optics had corrected for some of the perturbations.

The examples are given for telescope diameters of 1 and 1.8 m,
and Fried parameters of 10 and 20 cm. The wavelength of the stellar light
is in the J, H, or K band of atmospheric transmission,
where the fiber core diameter is tailored
to move the cutoff wavelength of the monomode regime to the edges
of these bands. 

\end{abstract}

\maketitle
\section{Scope}

One of the promises of monomode fiber optics is the removal of
corrugations across pupil planes (``cleaning'' of the beams) by symmetric
weighting of the electric field across the pupil
\cite{MennessonJOSA19,WallnerJOSA19,KeenMNRAS326,LabadieAA471}.
The transformation of phase screens is studied by computation of 
the overlap with the anticipated far-field distribution of the
fiber optics of the detector system \cite{ShaklanAO27}.
We demonstrate the negative impact of the growing number of speckles \cite{RoddierJopt14}
on the throughput of the spatial filter---represented by the fiber---as
a function of telescope diameter.
This combines essentially the work of Shellan \cite{ShellanJOSAA21}
---which describes the on-axis intensity after AO correction---
with the specific spatial filtering of a fiber, which is roughly of
Gaussian shape and therefore de-emphasizes the role of
higher modes in Zernike expansions.

\section{Optical Setup and Model} 

\subsection{Fibre Set} 

Step-index fibers are modeled with core radius $a$, refractive indices
$n_c$ in the core and $n$ in the cladding, and the same numerical aperture
\begin{equation}
\alpha_m=1.4 =\sqrt{n_c^2+n^2}
\end{equation}
for all bands. The difference in $n_c-n$ is kept at
0.36 percent---copied from a 
specification of the Corning SMF-28e Photonic Fiber.
This is rather hypothetical since we do not look at the
absorption characteristics of materials in the infrared \cite{LaurentPhd,DirnwoberDip}.
The cutoff wavelengths $\lambda_c$ considered here
are derived from the normalized frequency $v\approx 2.405$ \cite{GlogeAO10}
\begin{equation}
v =a k \alpha_m,
\end{equation}
where $k\equiv 2\pi/\lambda$ is the momentum number, which leads
to the fiber specifications of Table \ref{tab:fib}.

\begin{table}[hbt]
\caption{Choice of fiber geometries, cutoff wavelengths $\lambda_c$ and
core radii $a$ in infrared bands.
\label{tab:fib}
}
\label{tab.lambdac}
\begin{ruledtabular}
\begin{tabular}{c|cccc}
band & $\lambda_c$ ($\mu$m) & $a$ ($\mu$m) & $n_c$ & $n$  \\
\hline
J & $1.13$ & $3.089$ & 1.65437 & $1.64843$ \\
H & $1.36$ & $3.718$ & 1.65437 & $1.64843$ \\
K & $1.87$ & $5.112$ & 1.65437 & $1.64843$
\end{tabular}
\end{ruledtabular}
\end{table}

The two-dimensional distribution of the electric field
at the fiber's entrance
is computed by matching the two Bessel Functions in the core and in the
cladding at each individual wavelength $\lambda$ \cite{GlogeAO10,StolenAO14}.
The electric field functions $f(\rho)$
depend only on the distance $\rho$ to the fiber axis (and parametrically
on $\lambda$, $a$, $n_c$ and $n$), since no azimuthal
dependence is left in the monomode regime.
A numerical Hankel transform
\cite{AgnesiJOSA10,FerrariJOSAA16,MagniJOSA9,MarkhamJOSA20,PercianteJOSAA21,StolenAO14,CerjanJOSAA24}
transforms this into the far field $F({\bf r})$
\begin{equation}
F(r)\propto\int_0^\infty d\rho\,\rho f(\rho) J_0(\rho k\alpha)
\end{equation}
in the pupil plane.
(No Gaussian approximation to the far field \cite{RuilierSPIE3350,WagnerAO21}
is introduced.)
The integration over the azimuthal angle over the fiber's cut has already
been performed in the Fraunhofer approximation, and has been condensed
to the Bessel Function $J_0$. $r$ is the radial coordinate in the pupil plane,
$\alpha\le \alpha_m$ the angle from the fiber axis to this point
in the pupil plane. The factor $\rho$
is the Jacobian
from the introduction of circular coordinates in the plane of the fiber's
front face. The far field does not depend on the azimuthal angle
$\theta$ in the pupil plane.

\subsection{Phase Screens} 

The electric field in the exit pupil of the telescope is written
as a two-dimensional phase screen over the radial coordinate $r$
and azimuthal coordinate $\theta$ as
\begin{equation}
E({\bf r})=e^{i\varphi(r,\theta)}.
\label{eq:Eofr}
\end{equation}
Amplitude variations---scintillation as opposed to phase variations, or
imaging characteristics
\cite{KouznetsovAO36,MartinAO27,RodriguezAO44,GuyonAA387,StromqvistAO46}---are
not studied here,
so the phases $\varphi({\bf r})$ are kept real-valued.
Since we shall look only at coupling coefficients, the modulus of $E({\bf r})$
is arbitrarily normalized to unity.

The phase screens have been generated
with a Kolmogorov spectrum by synthesizing Karhunen-Lo\`eve (KL) basis functions
as described in the literature \cite{FriedJOSA68,WangJOSA68,RoddierOE29,MatharArxiv0705b}:
\begin{equation}
\varphi(r)=\sum_{p,q} g_{p,q}
K_p^{(q)}(r)
\Theta_q(\theta)
.
\end{equation}
We use normalized azimuthal basis functions
\begin{equation}
\Theta_q(\theta)\equiv \sqrt{\frac{\epsilon_q}{2\pi}}\times
\left\{ \begin{array}{c}
\cos q\theta\\
\sin q\theta
\end{array}\right. ,
\label{eq:thetanorm}
\end{equation}
where
\begin{equation}
\epsilon_q\equiv \left\{ \begin{array}{c@{,\quad}c}
1 & q=0\\
2 & q\ge 1
\end{array}\right.
\end{equation}
is Neumann's factor. If the radial basis functions are normalized
according to
\begin{equation}
\int_0^{D/2} r K_p^{(q)}(r)K_{p'}^{(q')}(r)dr = \delta_{pp'}
,
\end{equation}
the variance of the expansion coefficients is
\begin{equation}
\Var g_{p,q} = D^2 [D/r_0(\lambda)]^{5/3} {\cal B}_{p,q}^2,
\label{eq:varg}
\end{equation}
where ${\cal B}^2$ are the eigenvalues of a reduced KL equation \cite{FriedJOSA68}.
Note that there is some arbitrariness in distributing factors here:
the factor $D^2$  in (\ref{eq:varg}) might be absorbed in a renormalization
of $K_p^{(q)}$, and the factor $\surd{\epsilon_q/(2\pi)}$ of (\ref{eq:thetanorm})
could also be dispersed over $K_p^{(q)}$ and/or $\cal B$.

The turbulent atmosphere is represented by a Kolmogorov power-law
of the phase structure function---we do not discuss the validity of
this ansatz from any fundamental or experimental point of view.
The radial basis functions are generated numerically by solving the
symmetrized integral equations for the eigen-modes $K_p^{(q)}$;
Zernike polynomials \cite{XiJOSA24,DaiJOSAA24,SheppardAO43,ComastriJOpt9,BrummelaarOptComm132}
or polynomials fits
\cite{DaiJOSAA12}
have not been employed.

The Fried parameters $r_0$ that are quoted here are
those implicitly measured at $0.5$ $\mu$m and were actually scaled with
\cite{FriedJOSA68}
\begin{equation}
r_0(\lambda)=r_{0\vert 0.5\mu\mathrm{m}}
\left(\frac{\lambda}{0.5\mu\mathrm{m}}\right)^{6/5}
\label{eq:r0scale}
\end{equation}
to the infrared wavelength $\lambda$ to generate the kernel
of the KL integral equations.
The radial functions $K_p^{(q)}$ are re-generated for each
instance of the ratio $D/r_0(\lambda)$.

An Adaptive Optics (AO) correction parameter (degree) $c$ is introduced
which assumes that some set of low-order basis functions with
largest eigenvalues ${\cal B}_{p,q}$ is discarded
while building the full phase screen.
$c\ge 2$ means
tip and tilt are removed from each individual phase screen,
represented by the first line in \cite[Table III]{WangJOSA68};
$c\ge 5$ means correction through refocus and astigmatism and removal of
the first three lines in \cite[Table III]{WangJOSA68}, and $c=14$ assumes
correction of the modes of the first eight lines of
\cite[Table III]{WangJOSA68}.
The calculations were done on a set of
75 basis functions, sufficiently large in comparison to these low-order
corrections.

The survival of speckles is demonstrated by coupling 800 phase screens into a
virtual photometric channel as if one would project a single telescope's
input, directly with the
coupling lens onto the fiber head.

\subsection{Coupling} \label{sec:Coupl} 

The intensity coupling efficiency is computed from numerical evaluation
of overlap integrals over the circular pupil
\cite{ShaklanAO27,WagnerAO21},
\begin{equation}
A=\frac{|\int_{NA}E({\bf r})F(r)d^2r|^2}
{\int_{NA} E^*({\bf r})E({\bf r})d^2r\,\int_{NA}F^2(r)d^2r}
.
\label{eq:ceff}
\end{equation}

The phase screen samples are created by generation of independent
phase screens from uncorrelated Gaussian random numbers for the expansion coefficients $g_{p,q}$;
in that respect no time scale is needed to define the
transit time from one sample of the Kolmogorov statistics
to another \cite{TubbsAO44,KellererAAp461};
there is no such parameter as the wind velocity of Taylor screens
\cite{FuscoJoptA6}.

Geometric imperfections like fiber misalignment
\cite{ToyoshimaJOSAA23,WagnerAO21}
or splicing
are not incorporated, nor Fresnel reflection losses at the fiber front end
or a central obscuration (shadow of a secondary mirror) \cite{HouAO45}.

For the Unit Telescopes of the Very Large Telescope Interferometer
equipped with AO this has been studied in
Felkel's thesis \cite{FelkelDip}.

For weak turbulence, equation (\ref{eq:Eofr})
can be expanded into its Taylor series
$1+i\varphi-\varphi^2/2\cdots$. This decomposes the overlap integral
in the numerator of (\ref{eq:ceff}) into a constant (which
represents the limit of optimum coupling efficiency, $r_0\rightarrow \infty$),
linear terms of azimuthal modes proportional to $\sin\theta$ or $\cos\theta$
which do not contribute because the integral vanishes, plus quadratic terms
which are quadratic in $g_{p,q}{\cal B}^2_{p,q}$.
The assessment of Ruilier and Fried that
the energy coupling is represented by these terms
remains basically correct, although the smearing with $F(r)$
diminishes the influence of terms of larger radial nodal numbers $p$.

\section{Phase Screen Statistics} 
\begin{figure}
\includegraphics[width=0.5\textwidth]{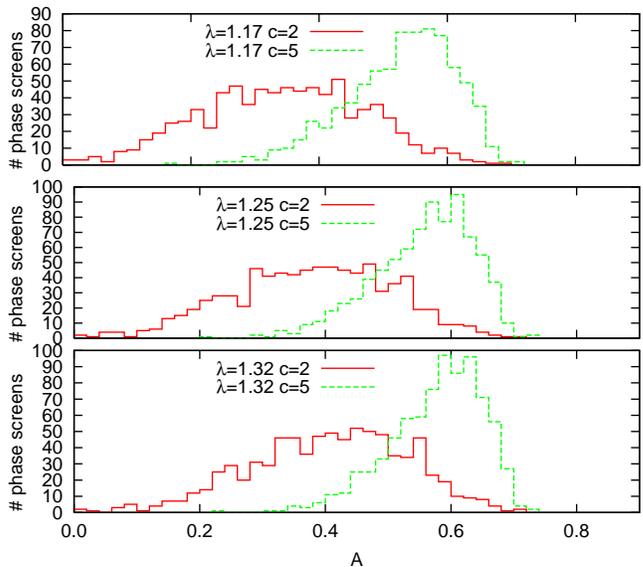}
\caption{
J-band.
$D=1$ m.
$r_0(500 nm)=0.1$ m.
\label{fig:fibKH_1_01_J.ps}
}
\end{figure}

\begin{figure}
\includegraphics[width=0.5\textwidth]{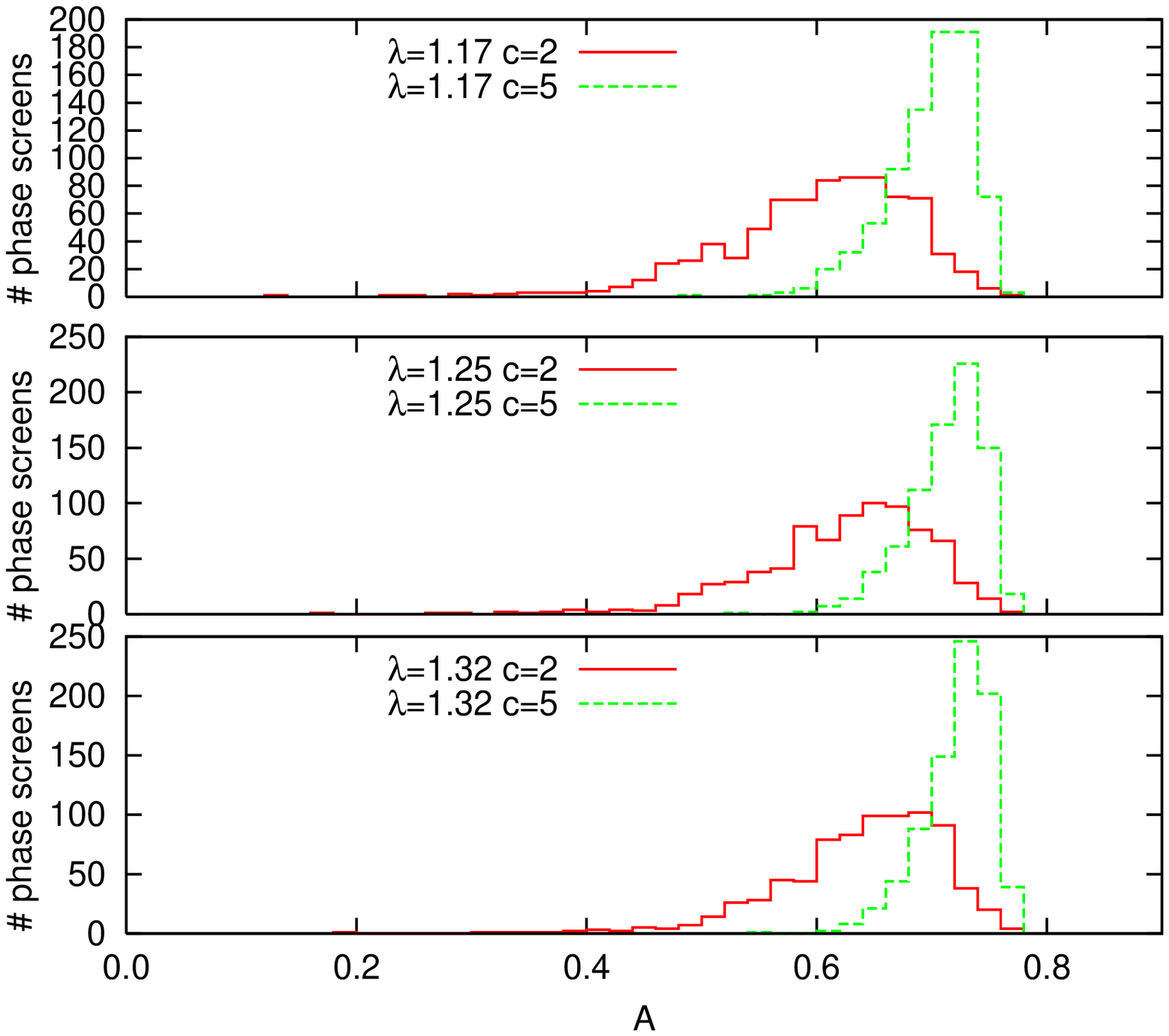}
\caption{
J-band.
$D=1$ m.
$r_0(500 nm)=0.2$ m.
\label{fig:fibKH_1_02_J.ps}
}
\end{figure}

\begin{figure}
\includegraphics[width=0.5\textwidth]{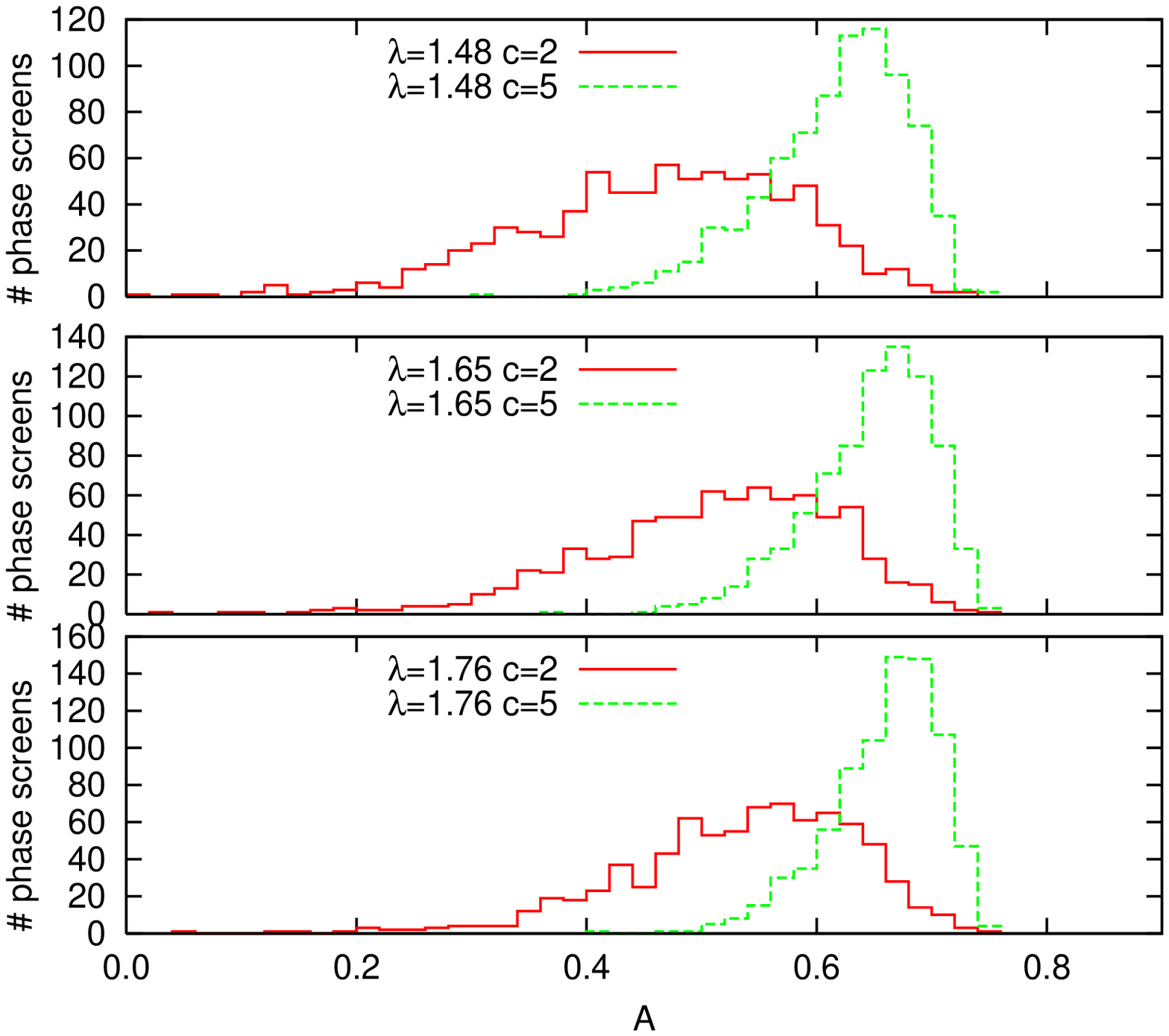}
\caption{
H-band.
$D=1$ m.
$r_0(500 nm)=0.1$ m.
\label{fig:fibKH_1_01_H.ps}
}
\end{figure}

\begin{figure}
\includegraphics[width=0.5\textwidth]{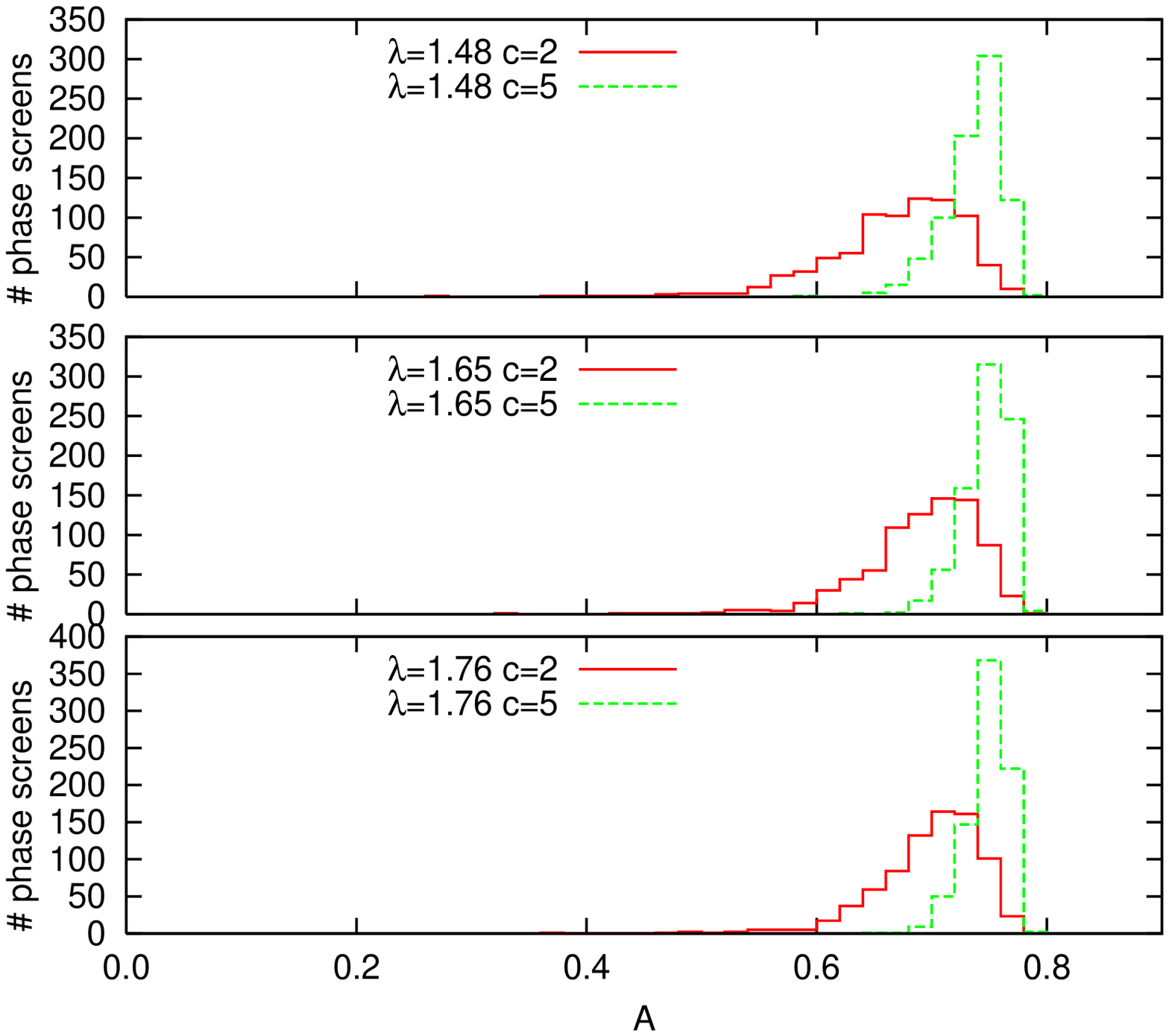}
\caption{
H-band.
$D=1$ m.
$r_0(500 nm)=0.2$ m.
\label{fig:fibKH_1_02_H.ps}
}
\end{figure}

\begin{figure}
\includegraphics[width=0.5\textwidth]{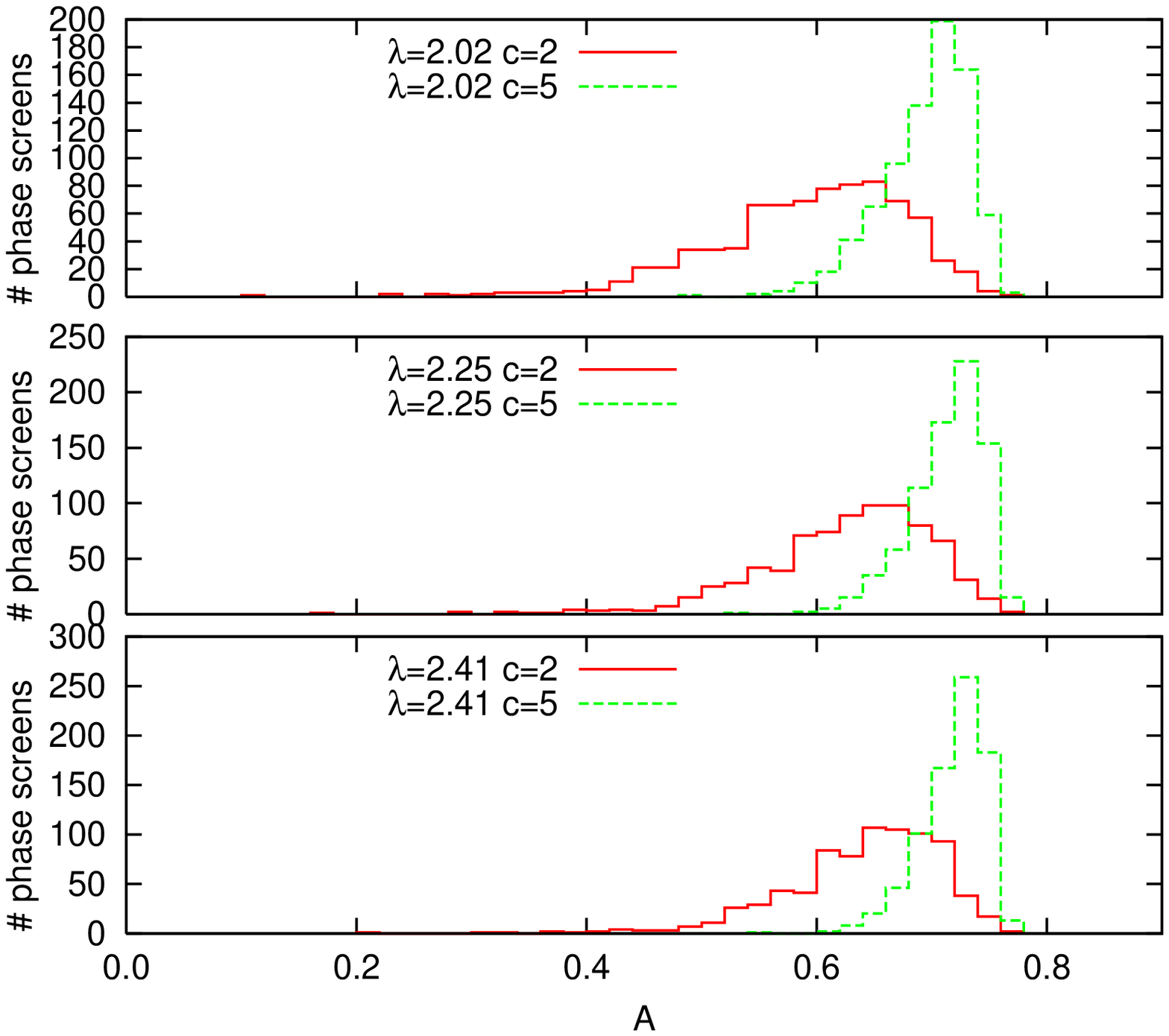}
\caption{
K-band.
$D=1$ m.
$r_0(500 nm)=0.1$ m.
\label{fig:fibKH_1_01_K.ps}
}
\end{figure}

\begin{figure}
\includegraphics[width=0.5\textwidth]{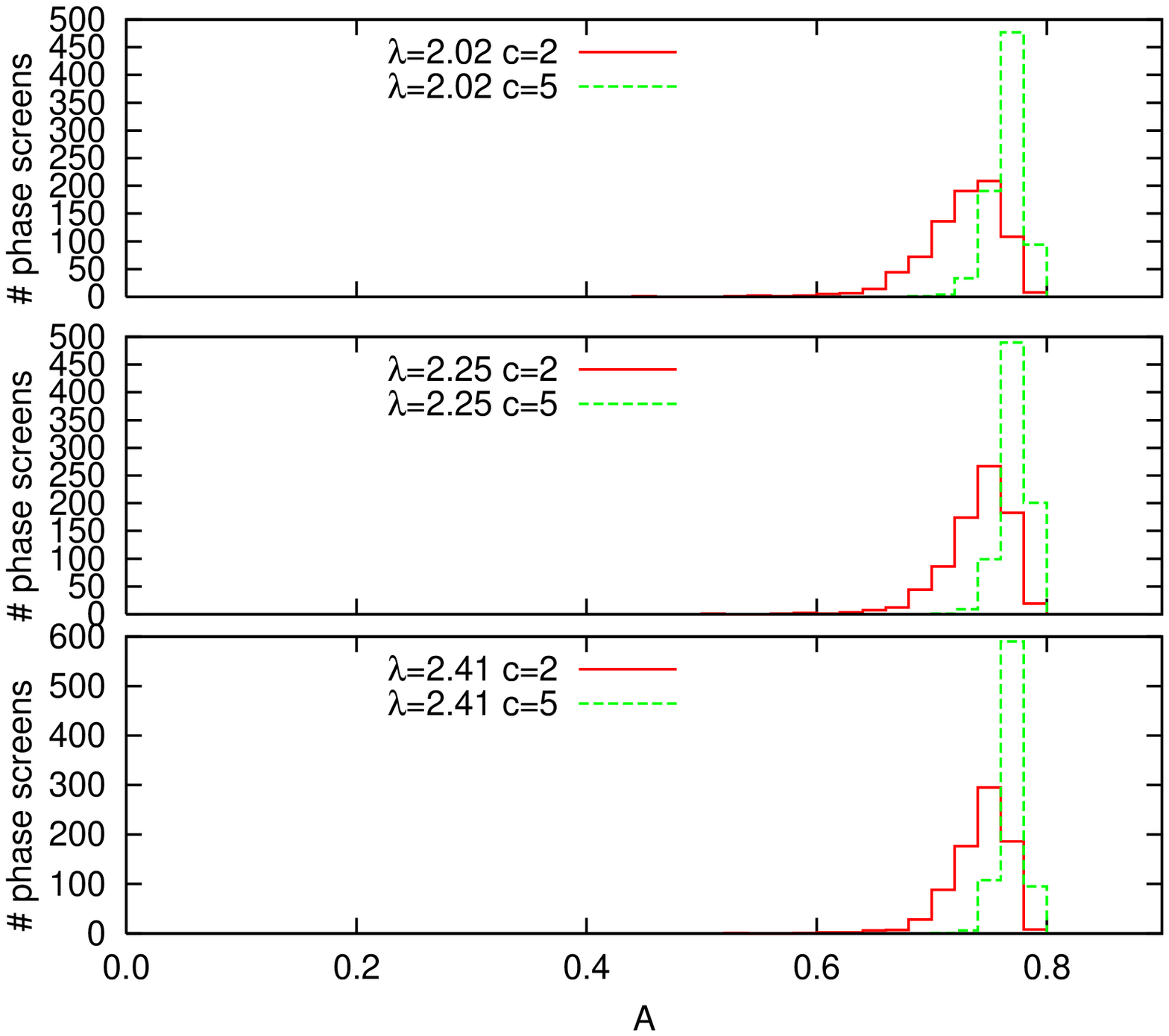}
\caption{K-band.
$D=1$ m.
$r_0(500 nm)=0.2$ m.
\label{fig:fibKH_1_02_K.ps}
}
\end{figure}

\begin{figure}
\includegraphics[width=0.5\textwidth]{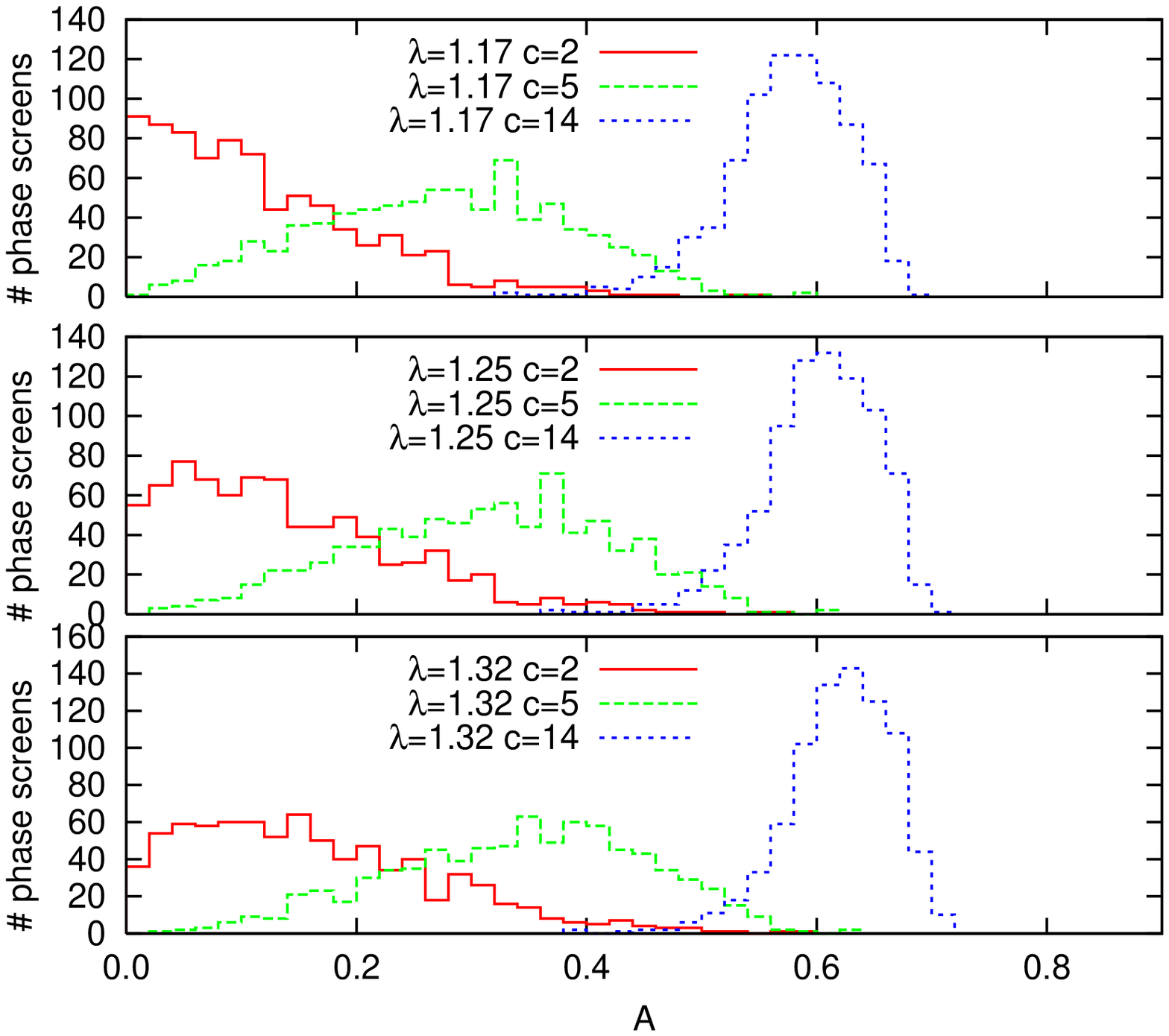}
\caption{
J-band.
$D=1.8$ m.
$r_0(500 nm)=0.1$ m.
\label{fig:fibKH_18_01_J.ps}
}
\end{figure}

\begin{figure}
\includegraphics[width=0.5\textwidth]{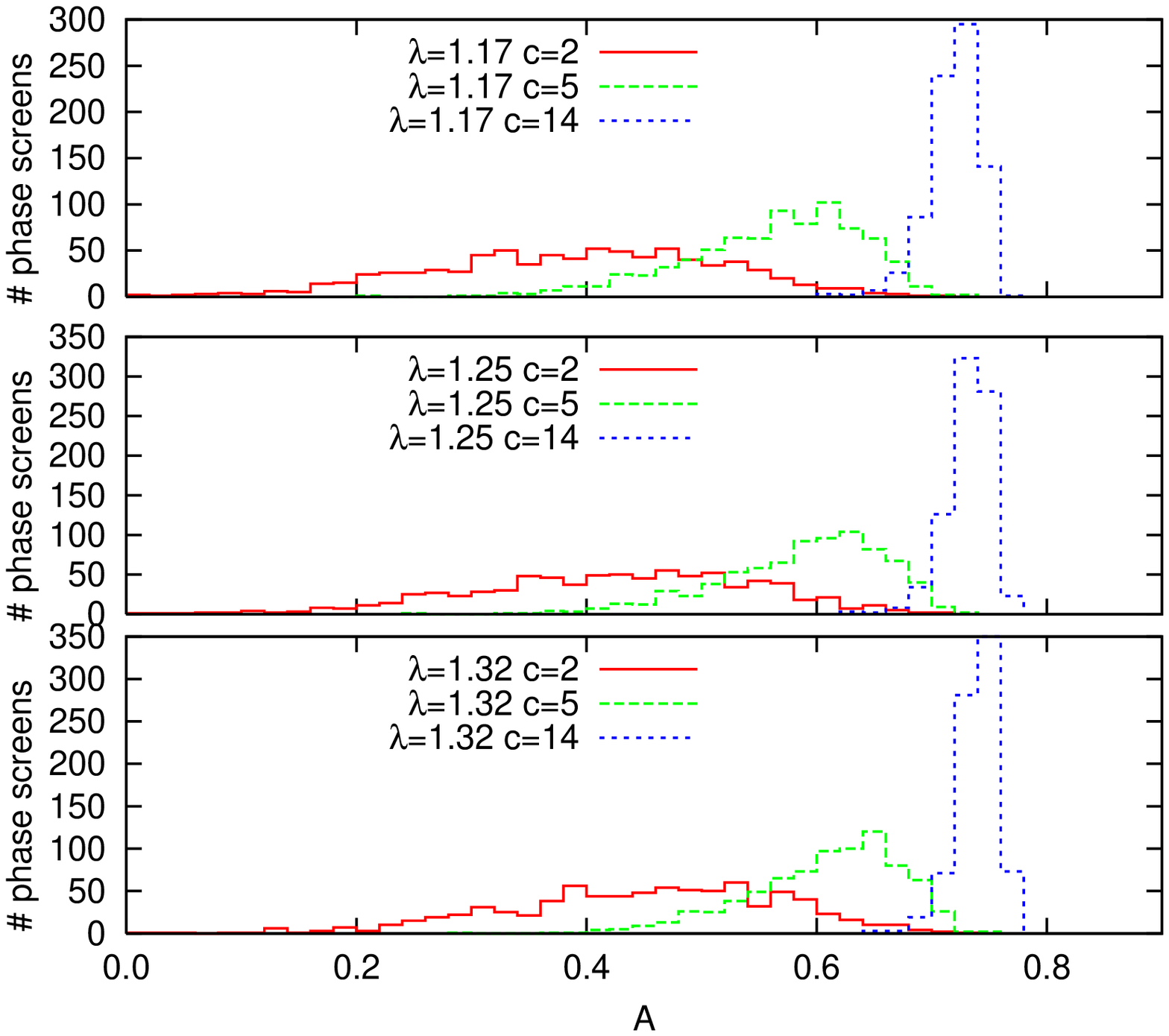}
\caption{
J-band.
$D=1.8$ m.
$r_0(500 nm)=0.2$ m.
\label{fig:fibKH_18_02_J.ps}
}
\end{figure}

\begin{figure}
\includegraphics[width=0.5\textwidth]{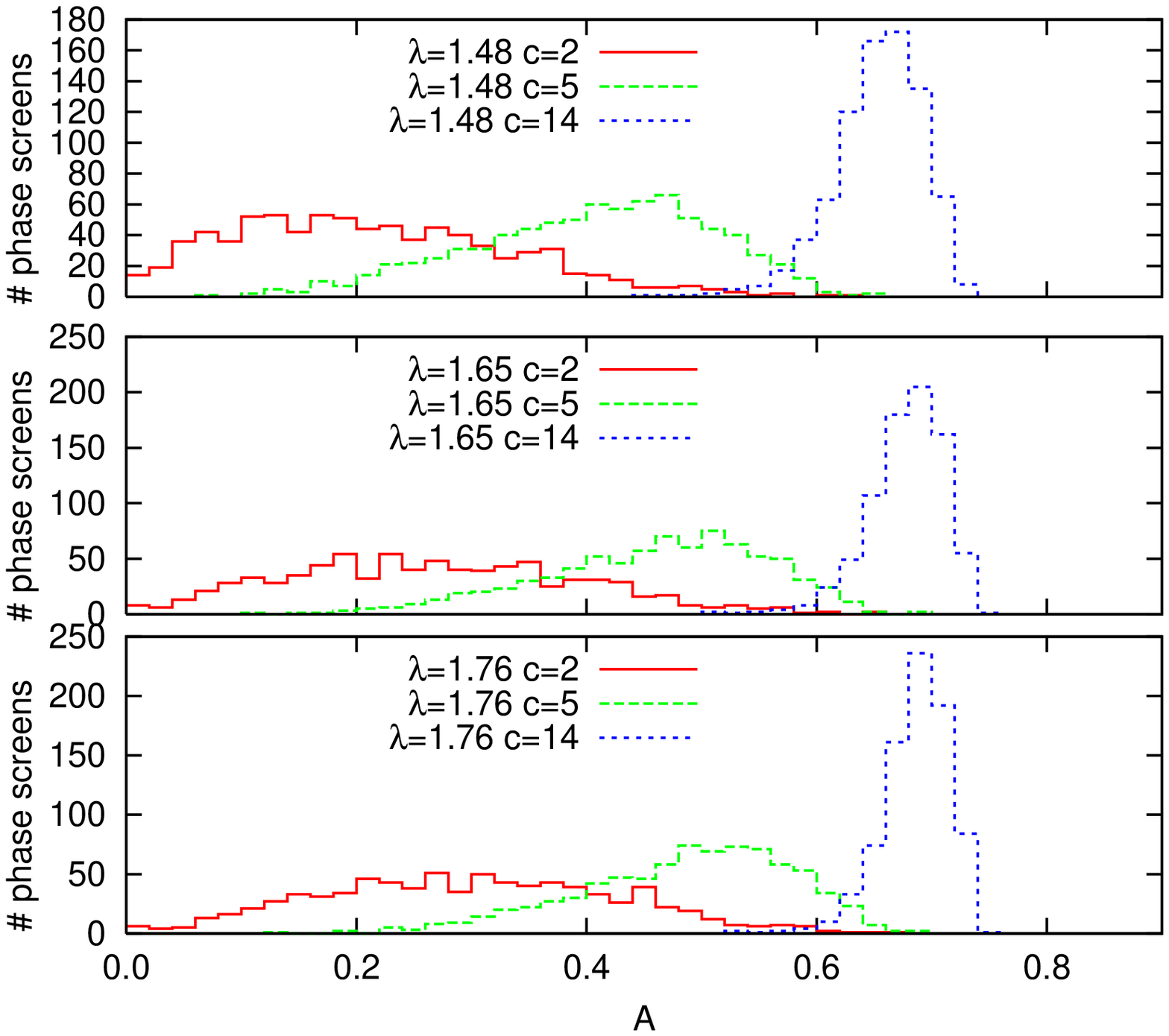}
\caption{
H-band.
$D=1.8$ m.
$r_0(500 nm)=0.1$ m.
\label{fig:fibKH_18_01_H.ps}
}
\end{figure}

\begin{figure}
\includegraphics[width=0.5\textwidth]{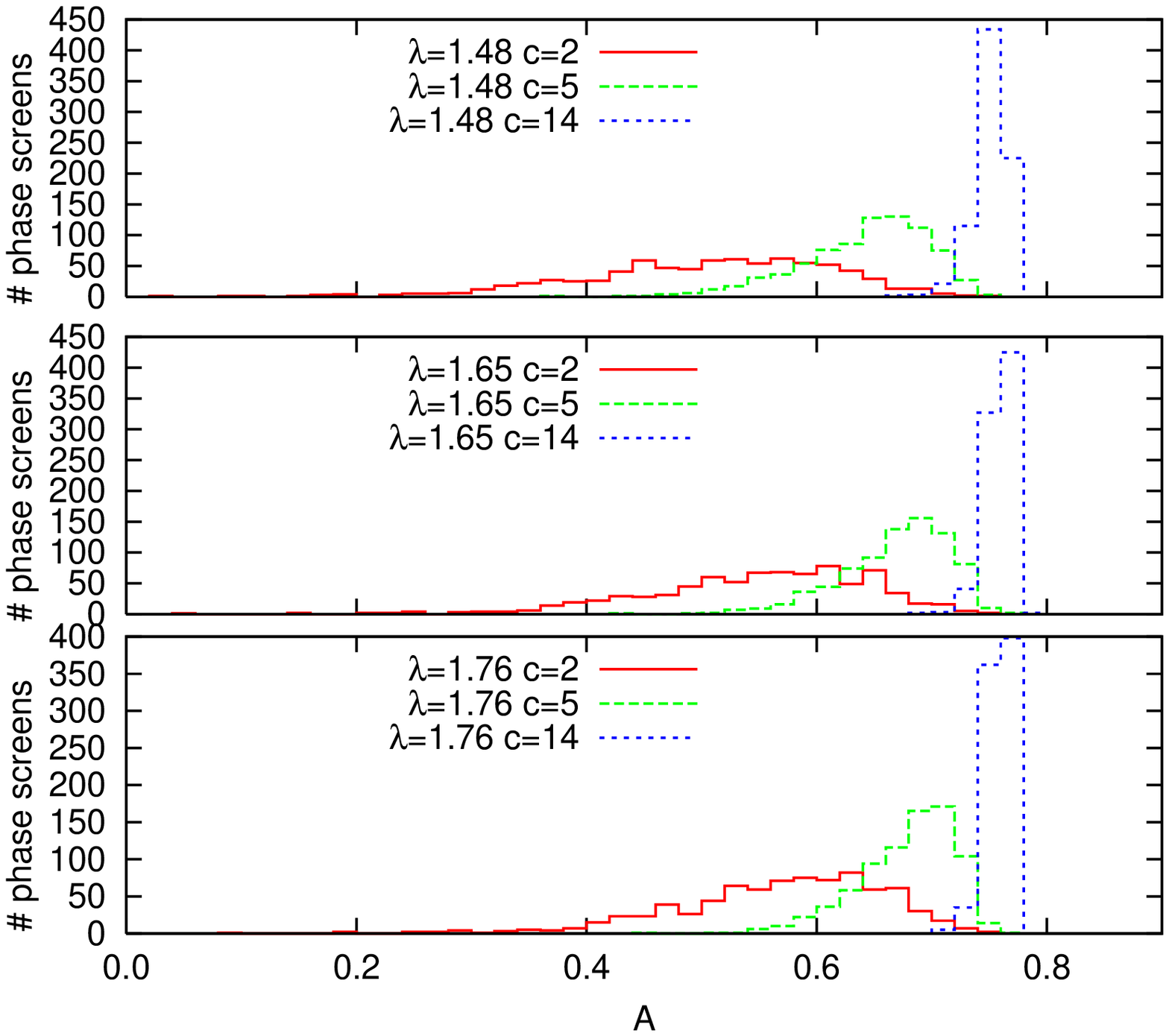}
\caption{
H-band.
$D=1.8$ m.
$r_0(500 nm)=0.2$ m.
\label{fig:fibKH_18_02_H.ps}
}
\end{figure}

\begin{figure}
\includegraphics[width=0.5\textwidth]{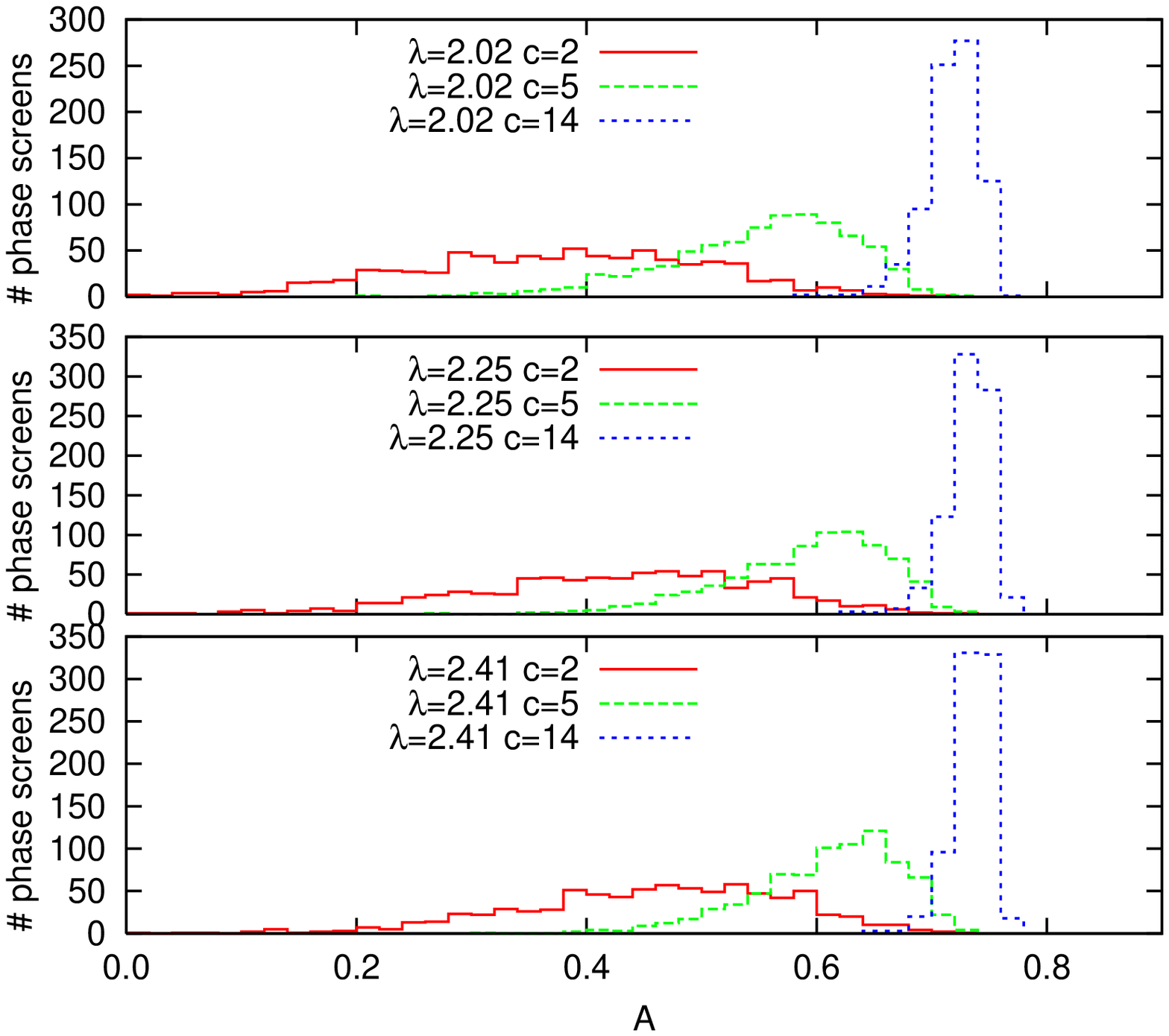}
\caption{
K-band.
$D=1.8$ m.
$r_0(500 nm)=0.1$ m.
\label{fig:fibKH_18_01_K.ps}
}
\end{figure}

\begin{figure}
\includegraphics[width=0.5\textwidth]{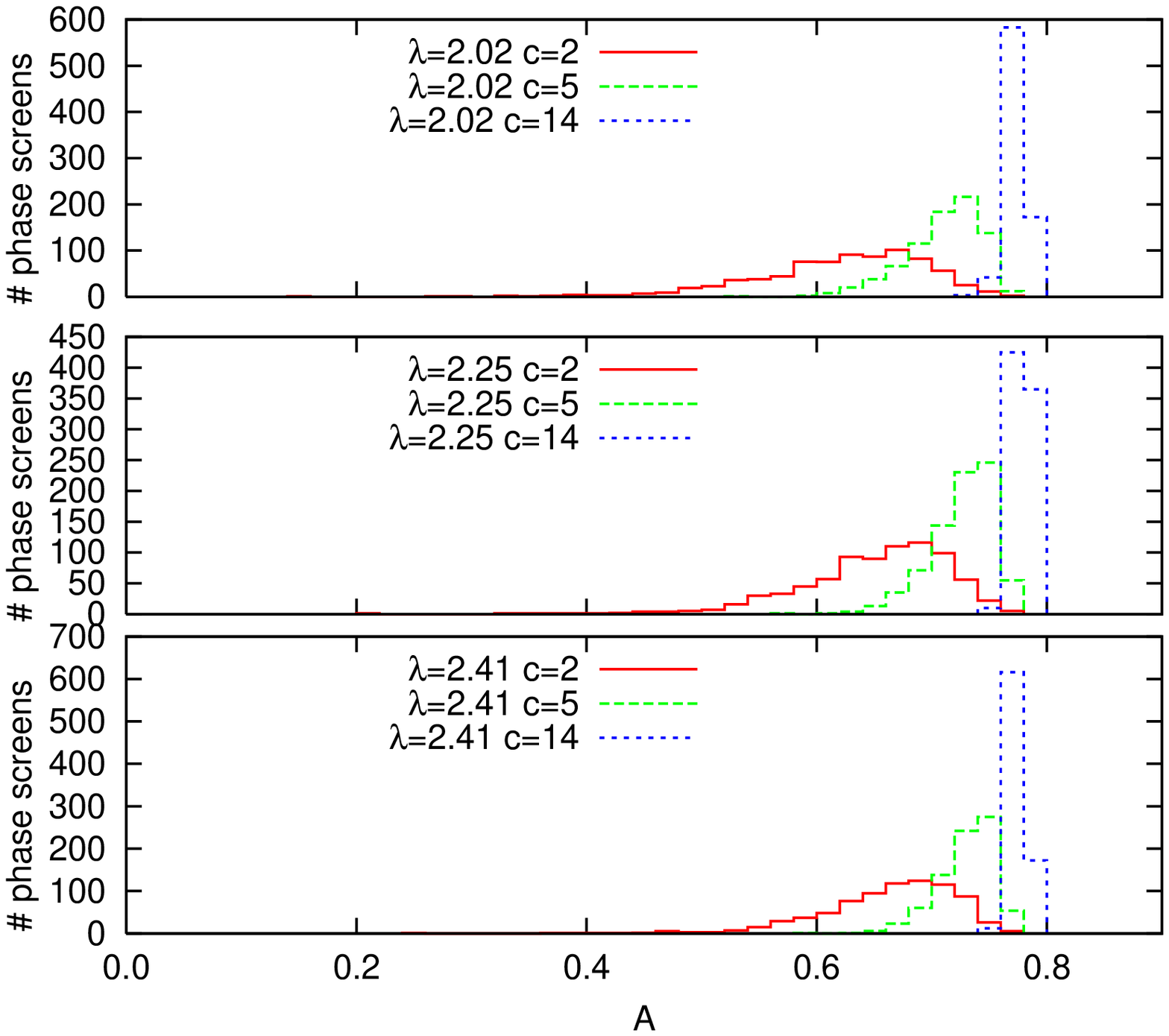}
\caption{K-band.
$D=1.8$ m.
$r_0(500 nm)=0.2$ m.
\label{fig:fibKH_18_02_K.ps}
}
\end{figure}

Statistics over 800 phase screens
have been compiled for sub-average ($r_0=10$ cm) and better-than-average 
($r_0=20$ cm) seeing conditions \cite{FuscoJoptA6,PercheronSPIE6268}
for telescopes of $D=1$ m (Figures \ref{fig:fibKH_1_01_J.ps}--\ref{fig:fibKH_1_02_K.ps})
and 1.8 m in diameter
(Figures \ref{fig:fibKH_18_01_J.ps}--\ref{fig:fibKH_18_02_K.ps})
at AO corrections levels of $c=2$, 5 and 14.

In each band, illumination by three different wavelengths $\lambda$
is studied, and the figures that follow are split into three
panels.
Standard diffraction theory shows how the \emph{blue} components of
spectra are enhanced for coupling into a waveguide of fixed
geometry \cite{PuJopt24}, but there is a counter-effect through
the smoothing of the phase structure functions with (\ref{eq:r0scale}).
Here, the net effect is a higher coupling efficiency for the \emph{red} end
of the bands, in accordance with earlier results \cite[Fig.\ 1]{ShaklanAO27}.
This is not necessarily the full truth since the phase
screen basis functions $K_p^{(q)}$ are calculated for each individual
wavelength as a function of its phase structure function; as our model leaves
a number $c$  of these aside, we are implicitly presuming that the AO
performs on the same level at all wavelengths within a band, and this
might not be realistic.

Tables \ref{tab:perc1} and \ref{tab:perc18} summarize
the cumulative distribution function of
each statistics by a triplet of values indicating the median (50 \% percentile)
and the offsets from there to the 84.1 \% and 15.9 \% percentiles,
equivalent to providing error bars on a 1$\sigma$ level. The notation is
$A_{50}+A_{84.1}-A_{15.9}$ where 50 percent of the coupling efficiencies
are smaller than $A_{50}$, 15.9 percent are smaller than $A_{50}-A_{15.9}$,
and 15.9 percent are larger than $A_{50}+A_{84.1}$.

\small
\begin{table}[hbt]
\caption{
Summary of percentiles of coupling efficiencies for 
a telescope diameter of $D=1$ m.
}
\label{tab:perc1}
\begin{ruledtabular}
\begin{tabular}{cccccc}
band & $r_0$ (500 nm) & Fig. & $\lambda$ & $c$ & ${A_{50}}^{+A_{84.1}}_{-A_{15.9}}$
\\
       & (m)  &  & ($\mu$m) & & \\
\hline
J & 0.1 & \ref{fig:fibKH_1_01_J.ps} & 1.17 & 2 &
$0.35^{+0.13}_{-0.14}$ \\
J & 0.1 & \ref{fig:fibKH_1_01_J.ps} & 1.17 & 5 &
$0.54^{+0.07}_{-0.10}$ \\
J & 0.1 & \ref{fig:fibKH_1_01_J.ps} & 1.25 & 2 &
$0.39^{+0.13}_{-0.14}$ \\
J & 0.1 & \ref{fig:fibKH_1_01_J.ps} & 1.25 & 5 &
$0.57^{+0.06}_{-0.09}$ \\
J & 0.1 & \ref{fig:fibKH_1_01_J.ps} & 1.32 & 2 &
$0.42^{+0.12}_{-0.14}$ \\
J & 0.1 & \ref{fig:fibKH_1_01_J.ps} & 1.32 & 5 &
$0.59^{+0.06}_{-0.09}$ \\
J & 0.2 & \ref{fig:fibKH_1_02_J.ps} & 1.17 & 2 &
$0.61^{+0.07}_{-0.10}$ \\
J & 0.2 & \ref{fig:fibKH_1_02_J.ps} & 1.17 & 5 &
$0.71^{+0.03}_{-0.04}$ \\
J & 0.2 & \ref{fig:fibKH_1_02_J.ps} & 1.25 & 2 &
$0.64^{+0.06}_{-0.09}$ \\
J & 0.2 & \ref{fig:fibKH_1_02_J.ps} & 1.25 & 5 &
$0.72^{+0.02}_{-0.04}$ \\
J & 0.2 & \ref{fig:fibKH_1_02_J.ps} & 1.32 & 2 &
$0.65^{+0.05}_{-0.08}$ \\
J & 0.2 & \ref{fig:fibKH_1_02_J.ps} & 1.32 & 5 &
$0.73^{+0.02}_{-0.03}$ \\
H & 0.1 & \ref{fig:fibKH_1_01_H.ps} & 1.48 & 2 &
$0.48^{+0.11}_{-0.14}$ \\
H & 0.1 & \ref{fig:fibKH_1_01_H.ps} & 1.48 & 5 &
$0.63^{+0.05}_{-0.07}$ \\
H & 0.1 & \ref{fig:fibKH_1_01_H.ps} & 1.65 & 2 &
$0.53^{+0.09}_{-0.12}$ \\
H & 0.1 & \ref{fig:fibKH_1_01_H.ps} & 1.65 & 5 &
$0.66^{+0.04}_{-0.06}$ \\
H & 0.1 & \ref{fig:fibKH_1_01_H.ps} & 1.76 & 2 &
$0.55^{+0.08}_{-0.12}$ \\
H & 0.1 & \ref{fig:fibKH_1_01_H.ps} & 1.76 & 5 &
$0.67^{+0.04}_{-0.06}$ \\
H & 0.2 & \ref{fig:fibKH_1_02_H.ps} & 1.48 & 2 &
$0.68^{+0.04}_{-0.07}$ \\
H & 0.2 & \ref{fig:fibKH_1_02_H.ps} & 1.48 & 5 &
$0.74^{+0.02}_{-0.03}$ \\
H & 0.2 & \ref{fig:fibKH_1_02_H.ps} & 1.65 & 2 &
$0.70^{+0.04}_{-0.06}$ \\
H & 0.2 & \ref{fig:fibKH_1_02_H.ps} & 1.65 & 5 &
$0.75^{+0.01}_{-0.02}$ \\
H & 0.2 & \ref{fig:fibKH_1_02_H.ps} & 1.76 & 2 &
$0.71^{+0.03}_{-0.05}$ \\
H & 0.2 & \ref{fig:fibKH_1_02_H.ps} & 1.76 & 5 &
$0.75^{+0.01}_{-0.02}$ \\
K & 0.1 & \ref{fig:fibKH_1_01_K.ps} & 2.02 & 2 &
$0.61^{+0.07}_{-0.10}$ \\
K & 0.1 & \ref{fig:fibKH_1_01_K.ps} & 2.02 & 5 &
$0.70^{+0.03}_{-0.05}$ \\
K & 0.1 & \ref{fig:fibKH_1_01_K.ps} & 2.25 & 2 &
$0.64^{+0.06}_{-0.09}$ \\
K & 0.1 & \ref{fig:fibKH_1_01_K.ps} & 2.25 & 5 &
$0.72^{+0.02}_{-0.04}$ \\
K & 0.1 & \ref{fig:fibKH_1_01_K.ps} & 2.41 & 2 &
$0.65^{+0.05}_{-0.08}$ \\
K & 0.1 & \ref{fig:fibKH_1_01_K.ps} & 2.41 & 5 &
$0.72^{+0.02}_{-0.03}$ \\
K & 0.2 & \ref{fig:fibKH_1_02_K.ps} & 2.02 & 2 &
$0.73^{+0.03}_{-0.04}$ \\
K & 0.2 & \ref{fig:fibKH_1_02_K.ps} & 2.02 & 5 &
$0.77^{+0.01}_{-0.02}$ \\
K & 0.2 & \ref{fig:fibKH_1_02_K.ps} & 2.25 & 2 &
$0.74^{+0.02}_{-0.03}$ \\
K & 0.2 & \ref{fig:fibKH_1_02_K.ps} & 2.25 & 5 &
$0.77^{+0.01}_{-0.01}$ \\
K & 0.2 & \ref{fig:fibKH_1_02_K.ps} & 2.41 & 2 &
$0.75^{+0.02}_{-0.03}$ \\
K & 0.2 & \ref{fig:fibKH_1_02_K.ps} & 2.41 & 5 &
$0.77^{+0.01}_{-0.01}$ \\
\end{tabular}
\end{ruledtabular}
\end{table}
\normalsize

\clearpage

\small
\begin{table}[hbt]
\caption{
Summary of percentiles of coupling efficiencies, $D=1.8$ m.
}
\label{tab:perc18}
\begin{ruledtabular}
\begin{tabular}{cccccc}
band & $r_0$ (500 nm) & Fig. & $\lambda$ & $c$ & ${A_{50}}^{+A_{84.1}}_{-A_{15.9}}$
\\
       & (m)  &  & ($\mu$m) & & \\
\hline
J & 0.1 & \ref{fig:fibKH_18_01_J.ps} & 1.17 & 5 &
$0.28^{+0.11}_{-0.12}$ \\
J & 0.1 & \ref{fig:fibKH_18_01_J.ps} & 1.17 & 14 &
$0.58^{+0.05}_{-0.06}$ \\
J & 0.1 & \ref{fig:fibKH_18_01_J.ps} & 1.25 & 5 &
$0.32^{+0.11}_{-0.12}$ \\
J & 0.1 & \ref{fig:fibKH_18_01_J.ps} & 1.25 & 14 &
$0.61^{+0.05}_{-0.05}$ \\
J & 0.1 & \ref{fig:fibKH_18_01_J.ps} & 1.32 & 5 &
$0.35^{+0.11}_{-0.12}$ \\
J & 0.1 & \ref{fig:fibKH_18_01_J.ps} & 1.32 & 14 &
$0.62^{+0.04}_{-0.05}$ \\
J & 0.2 & \ref{fig:fibKH_18_02_J.ps} & 1.17 & 5 &
$0.57^{+0.06}_{-0.09}$ \\
J & 0.2 & \ref{fig:fibKH_18_02_J.ps} & 1.17 & 14 &
$0.72^{+0.02}_{-0.02}$ \\
J & 0.2 & \ref{fig:fibKH_18_02_J.ps} & 1.25 & 5 &
$0.60^{+0.06}_{-0.08}$ \\
J & 0.2 & \ref{fig:fibKH_18_02_J.ps} & 1.25 & 14 &
$0.73^{+0.02}_{-0.02}$ \\
J & 0.2 & \ref{fig:fibKH_18_02_J.ps} & 1.32 & 5 &
$0.62^{+0.05}_{-0.08}$ \\
J & 0.2 & \ref{fig:fibKH_18_02_J.ps} & 1.32 & 14 &
$0.74^{+0.01}_{-0.02}$ \\
H & 0.1 & \ref{fig:fibKH_18_01_H.ps} & 1.48 & 5 &
$0.41^{+0.10}_{-0.12}$ \\
H & 0.1 & \ref{fig:fibKH_18_01_H.ps} & 1.48 & 14 &
$0.66^{+0.03}_{-0.04}$ \\
H & 0.1 & \ref{fig:fibKH_18_01_H.ps} & 1.65 & 5 &
$0.47^{+0.09}_{-0.11}$ \\
H & 0.1 & \ref{fig:fibKH_18_01_H.ps} & 1.65 & 14 &
$0.68^{+0.03}_{-0.03}$ \\
H & 0.1 & \ref{fig:fibKH_18_01_H.ps} & 1.76 & 5 &
$0.50^{+0.08}_{-0.11}$ \\
H & 0.1 & \ref{fig:fibKH_18_01_H.ps} & 1.76 & 14 &
$0.69^{+0.03}_{-0.03}$ \\
H & 0.2 & \ref{fig:fibKH_18_02_H.ps} & 1.48 & 5 &
$0.65^{+0.04}_{-0.06}$ \\
H & 0.2 & \ref{fig:fibKH_18_02_H.ps} & 1.48 & 14 &
$0.75^{+0.01}_{-0.01}$ \\
H & 0.2 & \ref{fig:fibKH_18_02_H.ps} & 1.65 & 5 &
$0.68^{+0.04}_{-0.05}$ \\
H & 0.2 & \ref{fig:fibKH_18_02_H.ps} & 1.65 & 14 &
$0.76^{+0.01}_{-0.01}$ \\
H & 0.2 & \ref{fig:fibKH_18_02_H.ps} & 1.76 & 5 &
$0.69^{+0.03}_{-0.05}$ \\
H & 0.2 & \ref{fig:fibKH_18_02_H.ps} & 1.76 & 14 &
$0.76^{+0.01}_{-0.01}$ \\
K & 0.1 & \ref{fig:fibKH_18_01_K.ps} & 2.02 & 5 &
$0.56^{+0.07}_{-0.09}$ \\
K & 0.1 & \ref{fig:fibKH_18_01_K.ps} & 2.02 & 14 &
$0.72^{+0.02}_{-0.02}$ \\
K & 0.1 & \ref{fig:fibKH_18_01_K.ps} & 2.25 & 5 &
$0.60^{+0.06}_{-0.08}$ \\
K & 0.1 & \ref{fig:fibKH_18_01_K.ps} & 2.25 & 14 &
$0.73^{+0.02}_{-0.02}$ \\
K & 0.1 & \ref{fig:fibKH_18_01_K.ps} & 2.41 & 5 &
$0.62^{+0.05}_{-0.07}$ \\
K & 0.1 & \ref{fig:fibKH_18_01_K.ps} & 2.41 & 14 &
$0.74^{+0.01}_{-0.02}$ \\
K & 0.2 & \ref{fig:fibKH_18_02_K.ps} & 2.02 & 5 &
$0.72^{+0.03}_{-0.04}$ \\
K & 0.2 & \ref{fig:fibKH_18_02_K.ps} & 2.02 & 14 &
$0.77^{+0.01}_{-0.01}$ \\
K & 0.2 & \ref{fig:fibKH_18_02_K.ps} & 2.25 & 5 &
$0.73^{+0.02}_{-0.03}$ \\
K & 0.2 & \ref{fig:fibKH_18_02_K.ps} & 2.25 & 14 &
$0.78^{+0.01}_{-0.01}$ \\
K & 0.2 & \ref{fig:fibKH_18_02_K.ps} & 2.41 & 5 &
$0.74^{+0.02}_{-0.03}$ \\
K & 0.2 & \ref{fig:fibKH_18_02_K.ps} & 2.41 & 14 &
$0.78^{+0.00}_{-0.01}$ \\
\end{tabular}
\end{ruledtabular}
\end{table}
\normalsize

\section{Summary}

\begin{figure}
\includegraphics[width=0.5\textwidth]{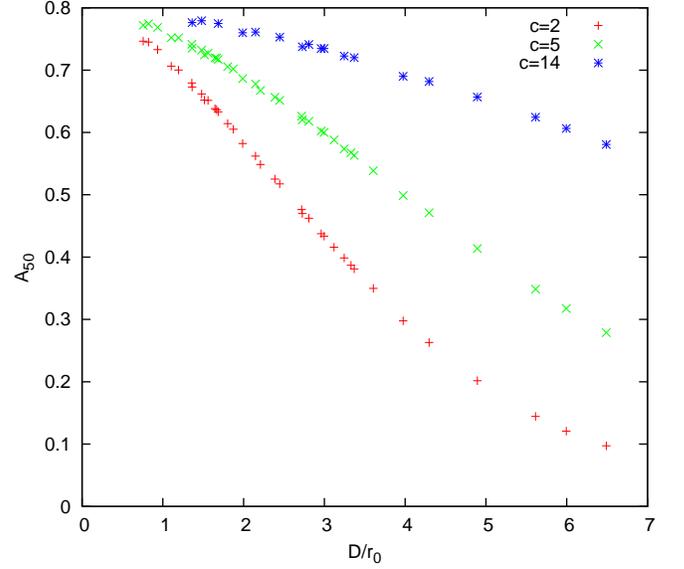}
\caption{
Coupling efficiencies (medians) sorted with respect to $D/r_0(\lambda)$
and the level of AO correction. Each curve of Figures \ref{fig:fibKH_1_01_J.ps}--\ref{fig:fibKH_18_02_K.ps}
is represented by one marker of its associated color.
\label{fig:fibKHstat.ps}
}
\end{figure}

A compact view on the median coupling efficiencies of these two tables is given
in Figure \ref{fig:fibKHstat.ps}\@.
The values on the curve with the red crosses, $c=2$,
are slightly more optimistic than those of Shaklan-Roddier
\cite[Fig. 2]{ShaklanAO27}
which one may attribute to underestimation of efficiencies
by the Gaussian approximation.

\begin{figure}
\includegraphics[width=0.5\textwidth]{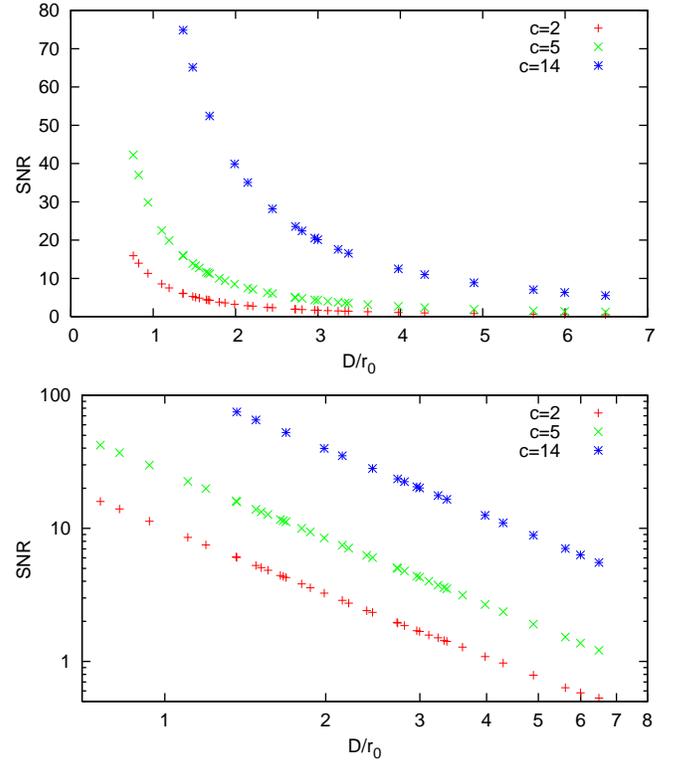}
\caption{
The signal-to-noise ratio represents the median coupling efficiency divided by
the statistical width of the distributions (1$\sigma$ error bars). Both plots
represent the same data set, one marker for each of the curves in
Figures \ref{fig:fibKH_1_01_J.ps}--\ref{fig:fibKH_18_02_K.ps}, the upper on linear and
the lower on logarithmic axis scales.
\label{fig:fibKHstatSNR.ps}
}
\end{figure}

By dividing the median through the noise introduced by the Kolmogorov
fluctuation in the phases, we obtain $A_{50}/(A_{84.1}-A_{15.9})$ as a
signal-to-noise ratio (SNR) for each combination of fiber geometry, wavelength,
telescope diameter, Fried parameter and AO correction.
Figure \ref{fig:fibKHstatSNR.ps} shows these twice.
The aim of
the doubly logarithmic
representation is to demonstrate that the SNR can be well fitted
by a power law $\propto (D/r_0)^{-5/3}$ with a prefactor depending
only on the AO correction level $c$ \cite{RuilierJOSA18}. This is not unexpected because
this power has been an input to equation (\ref{eq:varg}), and for a
pinhole type of spatial filter this remains essentially unharmed
as reasoned in Sect.\ \ref{sec:Coupl} \cite{NollJOSA66}.

Reduced coupling efficiencies as a function of the number of speckles
and degree of AO correction have been well reported by Shaklan and Roddier.
We have illustrated that in addition this reduction in photometric
signals is accompanied by wider variances of the expected
coupling efficiency, which leads to the need of longer integration
times
during observations.

\bibliographystyle{apsrmp}

\bibliography{all}

\end{document}